\documentclass[twocolumn,showpacs,aps,prd,superscriptaddress,amsmath,amssymb,floatfix]{revtex4}
\usepackage{graphicx}
\usepackage{bm}

\begin{document}
\title{Polarization effects in the Higgs boson decay to
$\gamma \, Z$ and test of $CP$ and $CPT$ symmetries}
\author{Alexander Yu. Korchin} \email{korchin@kipt.kharkov.ua}
\affiliation{NSC `Kharkov Institute of Physics and Technology',
61108 Kharkiv, Ukraine}
\affiliation{V.N.~Karazin Kharkiv National University, 61022 Kharkiv, Ukraine}
\author{Vladimir A. Kovalchuk}  \email{koval@kipt.kharkov.ua}
\affiliation{NSC `Kharkov Institute of Physics and Technology',
61108 Kharkiv, Ukraine}

\date{today}

\begin{abstract}
Polarization characteristics of $\gamma \gamma$ and $\gamma Z$
states in the Higgs boson decays $h \to \gamma \gamma$ and $h \to
\gamma Z$ are discussed. Based on effective Lagrangian, describing
$h \gamma \gamma $ and $h \gamma Z$ interactions with $CP$-even
and $CP$-odd parts, we calculate polarization parameters $\xi_1,
\, \xi_2, \, \xi_3$. A nonzero value of the photon circular
polarization, defined by parameter $\xi_2$, arises due to presence
of both parts in effective Lagrangian and its non-Hermiticity. The
circular polarization is proportional to the forward-backward
asymmetry of fermions in the decay $h \to \gamma \, Z \, \to
\gamma \, f\, \bar{f}$. Measurement of this observable would allow
one to search for deviation from the standard model and possible violation of
$CPT$ symmetry. We discuss also a possibility to measure
parameters $\xi_1, \, \xi_3$, describing correlation of linear
polarizations of photon and $Z$ boson, in the decay $h \to
\gamma^* \, Z \, \to \ell^+ \ell^- \, Z$ via distribution over the
azimuthal angle between the decay planes of $\gamma^* \to \ell^+
\ell^-$ and $Z \to \bar{f} f$. Deviation of the measured value of
$\xi_1$ from zero will indicate $CP$ violation in the Higgs
sector.

\end{abstract}


\maketitle

\setcounter{footnote}{0}

\section{\label{sec:Introduction}Introduction}

The ATLAS and CMS collaborations at the LHC have recently observed
\cite{ATLAS:2012,CMS:2012} a boson $h$ with mass around 126 GeV
with statistical significance of about five standard deviations.
The experimental evidence of this new particle is the strongest in
the two-photon and four-lepton final channels, where the detectors
give the best mass resolution.

Although the decay pattern of $h$ is mainly consistent with the
predictions of the standard model (SM), the clarification of the
nature of this particle still needs more data and time.
The spin of this boson is known to be zero or two, while the $CP$
properties are not yet ascertained. Recent data are more
consistent with the pure scalar boson hypothesis than the pure
pseudoscalar one~\cite{CMS:2012sp}. Though in the SM the Higgs
boson has $J^{PC} = 0^{++}$, there are many extensions of the SM
with a more complicated Higgs sector, in which some of the Higgs
bosons may not have definite $CP$ parity
\cite{Pilaftsis:1999np,Barger:2009pr,Branco:2012pre}.

This aspect of the Higgs study is also related to the origin of
the $CP$ violation. In the SM the source of the
$CP$ violation is the complex irreducible phase in the
Cabibbo-Kobayashi-Maskawa (CKM) matrix \cite{CKM}, however this
effect is not sufficient to explain the observed matter-antimatter
asymmetry in the Universe \cite{Bailin}. There may be other
mechanisms of the $CP$ violation beyond the CKM matrix, for
example, in the Higgs sector. From this point of view, the
elucidation of the $CP$ properties of the observed $h$ boson would
be an important step towards clarification of the mechanisms
giving rise to the masses of particles, their mixing and $CP$
violation.

Recently the $CP$ properties of the Higgs boson in the two-photon
decay channel $h \to \gamma \, \gamma$ have been addressed in
Ref.~\cite{Voloshin:2012}. In this channel the branching fraction, measured by the ATLAS collaboration, is larger than the value predicted
in the SM by a factor of $1.60 \pm 0.30 $ for $m_h=125.2 \pm
0.26\,{\rm (stat)}^{+0.5}_{-0.6} \, {\rm (syst)}\,{\rm GeV}$
\cite{ATLAS:2013_CONF034}, while the CMS collaboration obtained for this factor
$0.77\pm 0.27$ for $m_h=125.7\, \pm 0.3 {\rm (stat)} \pm 0.3 \,
{\rm (syst)}\,{\rm GeV} $ \cite{CMS:2013_HIG005}.
The author of \cite{Voloshin:2012}, in framework of a model with
vectorlike fermions, showed that the $CP$ violation in the $h \to
\gamma \, \gamma$ decay results in the dependence of the
differential decay rate on the angle between linear polarization
vectors of the photons. Experimentally, this angular distribution
can be measured after both photons are converted into the $e^+,
\,e^-$ pairs via the azimuthal angle distribution between the
planes spanned by the two $e^+, \, e^-$ pairs. In
Ref.~\cite{Kovalchuk:2008} a model-independent analysis of the
$CP$ violation effects in the Higgs boson into a pair of the gauge
bosons $W^+, \,W^-$ or $Z, \,Z$ has been presented. The author has
studied the angular distributions of the fermions $f = \ell, \, q$
in the cascade processes $ h \to V_1 \, V_2 \, \to (f_1 \,
\bar{f}_2) \, (f_3 \, \bar{f}_4)$ and analyzed possibilities of
observation of the $CP$ violation in these decays to various final
lepton and quark pairs.

In the present paper we would like to address the decay of the
Higgs boson to the photon and $Z$ boson, $h \to \gamma\, Z$,
pointing out to a possibility of studying in this decay not only
the $CP$ properties of the newly discovered boson, but also the
validity of the $CPT$ symmetry. In this connection one can recall
Ref.~\cite{Okun:2002} in which the author showed that an observation
of the circular polarization of the photon in the neutral pion
decay $\pi^0 \to \gamma \, \gamma$ (or $\eta \to \gamma \,
\gamma$)  would signal violation of the $CPT$ symmetry. Indeed,
the product $\vec{s} \, \vec{k} $ (where $\vec{s}$ \ is the photon
spin and $\vec{k}$ \ is its momentum) is $P$ odd and $T$ even.
Such a correlation in the $\pi^0$ decay arises due to interference
of the two terms in the interaction Lagrangian: a scalar
$\tilde{c}\, \pi^0 \, \epsilon^{\mu \nu \rho \sigma} \, F_{\mu
\nu} \, F_{\rho \sigma}$ and a pseudoscalar $c \, \pi^0 \, F_{\mu
\nu} \, F^{\mu \nu}$, with $\tilde{c}$ and $c$ being couplings
constants and $F_{\mu \nu} =
\partial_{\mu} A_{\nu} -\partial_{\nu} A_{\mu}$. The analysis of
\cite{Okun:2002} demonstrated that a nonzero value of $\vec{s} \,
\vec{k} $ correlation may appear due to a non-Hermiticity of the
tree-level amplitude, i.e. ${\rm Im} \tilde{c} \ne 0$ or/and ${\rm
Im} c \ne 0$, and/or higher-order loop corrections to the
amplitude inducing imaginary part of $\tilde{c}$.

Note that such a correlation in the Higgs boson decay to two
transversally polarized $Z$ bosons in connection with possible
violation of $CPT$ symmetry has been discussed in
\cite{Kovalchuk:2008}.

Generally, similar arguments can be applied to the two-photon
decay of the Higgs boson with an analogous conclusion. However
measurement of the photon circular polarization in the $h \to
\gamma \, \gamma$ decay is a rather difficult task. In the present
paper we suggest to study $CP$ and possible $CPT$ violation in
the decay
\begin{equation}\label{eq:001}
h \to \gamma \, Z \,  \to \gamma \, f\, \bar{f} \, ,
\end{equation}
with $f = \ell, \, q$. It turns out that the decay distribution
over the angle $\theta$ between the momentum of the fermion $f$
(in the rest frame of the $Z$) and momentum of the $Z$ (in the
rest frame of the $h$) gives information on the photon circular
polarization. Namely, a nonzero photon circular polarization
induces a term $\sim \cos \theta $ in this distribution which can
be measured through the forward-backward asymmetry $A_{\rm FB}$.

In the SM the $h \to \gamma \, Z$ decay amplitude in the lowest
order is determined by the loop contributions
\cite{Cahn:1979pl,Bergstrom:1985np} which have a small but nonzero
imaginary part arising due to rescattering effects $h \to f
\bar{f} \to \gamma \, Z$ for the fermions $f$ with masses $m_f \le
m_h/2$. The corresponding effective Lagrangian ${\cal
L}_{\rm eff}^{h\gamma Z}$, describing interaction of $h, \, \gamma$ and $Z$, is
thus non-Hermitian. Non-Hermiticity of effective Lagrangian leads to a
nonzero value of the net photon helicity once we assume a mixture of $CP$
violating term in ${\cal L}_{\rm eff}^{h\gamma Z}$.
Note that in the SM and theories beyond the SM which are $CPT$ symmetric,
there are no sources of non-Hermiticity of ${\cal L}_{\rm eff}^{h\gamma Z}$
apart from rescattering effects.

The $CPT$ theorem is one of the most profound results of quantum
field theory \cite{Streater:1964}. It is a consequence of Lorentz
invariance, locality, connection between spin and statistics, and
a Hermitian Hamiltonian. However there are many extensions of the
SM in which $CPT$ violation appears due to nonlocality in the
string theory, or violation of Lorentz symmetry in the extra
dimensional models (see, for example, \cite{Kostelecky:2008}). One
can also mention possible deviations from the standard quantum
mechanical evolution of states in some models of quantum gravity,
and the corresponding breakdown of the $CPT$ symmetry is investigated
in the neutral-meson system, where novel $CPT$-violating
observables for the $\phi$-factories and $B$-factories are
proposed~\cite{Bernabeu:2004}.
The $CPT$ violating effects in some of these underlying theories, in principle, can be additional sources of non-Hermiticity of effective Lagrangian ${\cal L}_{\rm eff}^{h\gamma Z}$
and hence contribute to photon circular polarization.

As for experimental results on the SM Higgs boson decay to the $Z$
boson and photon, we mention recent ATLAS and CMS
results ~\cite{ATLAS:2013_CONF009,CMS:2013}. The Higgs production
cross section times the $h\to \gamma\,Z$ branching fraction limits
are about an order of magnitude larger than the SM expectation for
$m_h=125$ GeV.

The paper is organized as follows. In Sec.~\ref{sec:formalism}
effective Lagrangian for $h\,\gamma\,\gamma$ and $h\, \gamma
\, Z$ interactions and coupling constants in the SM and some its
extensions are considered. In Sec.~\ref{sec:amplitudes}
amplitudes  and polarization parameters for the decays $h \to \gamma\,\gamma$ and $h \to \gamma \, Z$ are specified. Distribution of the $h \to \gamma
\, Z \,  \to \gamma \, f\, \bar{f}$ decay in the polar angle, and distribution
of the $h \to \gamma^* \, Z \, \to \ell^+ \ell^- \, Z$ decay (with $Z \to \bar{f} f$ on mass shell) in the azimuthal angle are obtained.
In Sec.~\ref{sec:results} results of calculation
and discussion are presented. In Sec.~\ref{sec:conclusions} we draw conclusions.


\section{\label{sec:formalism} Formalism }

The effective Lagrangian for the $h\,\gamma\,\gamma$ and $h\,
\gamma \, Z$ interactions can be written, as
\begin{equation}\label{eq:002}
{\cal L}_{\rm eff}^{h\gamma\gamma}=\frac{e^2}{32\,\pi^2
\,v}\left(c_\gamma\,F_{\mu\nu}F^{\mu\nu}h-{\tilde
c}_\gamma\,F_{\mu\nu}{\widetilde F}^{\mu\nu}h\right)\,,
\end{equation}
\begin{eqnarray}\label{eq:003}
&&{\cal L}_{\rm eff}^{h\gamma Z}=\frac{e\,g}{16\,\pi^2
\,v}\Bigl(c_{1Z}\,Z_{\mu\nu}F^{\mu\nu}h \nonumber \\
&-&c_{2Z}\left(\partial_\mu h\,Z_\nu-\partial_\nu
h\,Z_\mu\right)F^{\mu\nu}-{\tilde c}_Z\,Z_{\mu\nu}{\widetilde
F}^{\mu\nu}h\Bigr)\,,
\end{eqnarray}
where $e$ is the positron electric charge, $g$ is the $SU(2)_L$
coupling constant and $v=\left(\sqrt{2}G_{\rm F}\right)^{-1/2}
\approx 246$ GeV is the vacuum expectation value of the Higgs
field. Here $F_{\mu\nu}$ and $Z_{\mu\nu}$ are the standard field
strengths for the electromagnetic and $Z$ field and ${\widetilde
F}_{\mu\nu}=\varepsilon_{\mu\nu\alpha\beta}F^{\alpha\beta}/2$,
with convention $\varepsilon_{0123}=+1$. Dimensionless parameters
$c_\gamma$, $c_{1Z}$, $c_{2Z}$, ${\tilde c}_\gamma$, and ${\tilde
c}_Z$ are effective coupling constants~\cite{footnote1}.
As these coupling constants are, in general, complex-valued, the operators (\ref{eq:002}) and (\ref{eq:003}) are non-Hermitian, while being local and Lorentz invariant.

It is convenient to write
the couplings $c_\gamma$ and $c_{1Z}$ as the sums of terms in the
SM and new physics (NP) beyond the SM: \ $c_\gamma = c_\gamma^{\rm
SM} + c_\gamma^{\rm NP}$, \ $c_{1Z} = c_Z^{\rm SM} + c_{1Z}^{\rm
NP}$. In the SM, ${\tilde c}_\gamma=c_{2Z}={\tilde c}_Z=0$ and
their nonzero values come from effects of the NP. The couplings $c_\gamma^{\rm SM}$ and $c_Z^{\rm SM}$ have
small imaginary parts which arise due to the intermediate
on mass shell $\ell^+ \, \ell^- $ and $q \bar q$ \ states in the one-loop contributions
[where $\ell = e, \, \mu, \, \tau$ denote leptons and $q = u, \,
d, \, s, \, c, \, b \,$ denote quarks (excluding $t$ quark)]. In the one-loop
order $c_\gamma^{\rm SM}$ and $c_Z^{\rm SM}$ are given
by~\cite{Bergstrom:1985np,Spira:1998,Manohar:2006pl}
\begin{eqnarray}\label{eq:004}
c_\gamma^{\rm SM}&=&A_1^\gamma(\tau_W)+\sum_{f}
N_f\,Q_f^2\,A_{1/2}^\gamma(\tau_f)\nonumber \\
&\approx&-6.60+0.08 i \,,
\end{eqnarray}
\begin{eqnarray}\label{eq:005}
c_Z^{\rm SM}&=&-A_1^Z(\tau_W,\lambda_W)-\sum_{f}
N_f\,Q_f\,g_f\,A_{1/2}^Z(\tau_f\,,\lambda_f)\nonumber \\
&\approx& -5.540+0.005 i \,,
\end{eqnarray}
where $f= (\ell, \, q, \, t)$,\  $N_f = 1 \, (3)$ for leptons
(quarks), $Q_f$ is the charge of the fermion $f$ in units of the
electric charge of the positron. Here also $g_f=(2\,t_{3L, \, f}
-4 \, Q_f \sin^2 \theta_W)/\cos \theta_W$, where $t_{3L, \, f}$ is
the projection of the weak isospin of the $f$ fermion, and
$\theta_W$ is the weak angle. The one-loop functions $A_1^\gamma,
\, A_{1/2}^\gamma, \, A_1^Z, \, A_{1/2}^Z  $ are defined in the Appendix~\ref{sec:appendix}.
These functions depend on arguments
$\tau_W=4m^2_W/m^2_h$, \ $\lambda_W=4m^2_W/m^2_Z$, \
$\tau_f=4m^2_f/m^2_h$, \ $\lambda_f=4m^2_f/m^2_Z$, with $m_h$
being the mass of the Higgs boson, $m_W \, (m_Z)$ being the mass of the
$W$ ($Z$) boson, and $m_f$ being the mass of the $f$-th fermion. Numerical
values in (\ref{eq:004}), (\ref{eq:005}) are obtained for
$m_h=126$ GeV using the SM parameters from \cite{PDG:2012}, and
the quark masses are chosen according to \cite{Dittmaier:2011}.

The terms $c_\gamma$, $c_{1Z}$, and $c_{2Z}$ above
correspond to a $CP$-even scalar $h$, while the terms ${\tilde
c}_\gamma$ and ${\tilde c}_Z$ indicate a $CP$-odd pseudoscalar
$h$. The presence of both sets of terms means that $h$ is not a
$CP$ eigenstate. Interference of these terms lead to $CP$
violating effects which reveal in polarization states of the
photon. Generally, the couplings $c_\gamma^{\rm NP}, \,
c_{1Z}^{\rm NP},$ \ ${\tilde c}_\gamma$, \ $c_{2Z}$, \ ${\tilde
c}_Z$ may be complex.

The SM can be considered as effective low-energy theory of
an underlying unknown theory at a scale $\Lambda$ (characteristic
scale of the NP) which is much higher than the electroweak scale
$v$. In effective field-theory language
\cite{Manohar:2006pl,Buchmuller:1986np,Hagiwara:1993pl,Hagiwara:1993pr,Grzadkowski:2010jh,Grojean:2013},
the couplings $c_\gamma^{\rm NP}, \, c_{1Z}^{\rm NP},$ \ ${\tilde
c}_\gamma$, \ $c_{2Z}$, \ ${\tilde c}_Z$ can be obtained from
gauge invariant dimension-6 operators such as
\begin{eqnarray}\label{eq:006}
\mathcal{O}_{B} &=& i \frac{g^\prime}{\Lambda^2} \,
(D_\mu\,H)^\dagger \, (D_\nu\,H) \, B^{\mu\, \nu},\nonumber \\
\mathcal{O}_{W} &=& i \frac{g}{\Lambda^2} \,
(D_\mu\,H)^\dagger \,\tau_k (D_\nu\,H) \, W_k^{\mu\, \nu}, \nonumber \\
\mathcal{O}_{BB} &=&  \frac{\,{g^\prime}^2}{2 \, \Lambda^2} \,
H^\dagger \, H \, B_{\mu\, \nu} B^{\mu \, \nu}, \nonumber \\
\widetilde{\mathcal{O}}_{BB} &=&  \frac{\,{g^\prime}^2}{2 \, \Lambda^2} \, H^\dagger \,  H \,
{B}_{\mu\, \nu} \widetilde B^{\mu \, \nu}, \nonumber \\
\mathcal{O}_{WW} &=&  \frac{g^2}{2 \, \Lambda^2} \, H^\dagger \, H
\, W_{k\,\mu\, \nu} W_k^{\mu \, \nu}, \nonumber \\
\widetilde{\mathcal{O}}_{WW} &=&  \frac{g^2}{2 \, \Lambda^2} \, H^\dagger \,  H \, W_{k\,\mu\, \nu}
 \widetilde W_k^{\mu \, \nu}, \nonumber \\
\mathcal{O}_{WB} &=&  \frac{g^\prime \, g}{2 \, \Lambda^2} \,
H^\dagger \, \tau_k \, H \, W_k^{\mu \, \nu} B_{\mu\, \nu},\nonumber \\
\widetilde{\mathcal{O}}_{WB} &=&  \frac{g^\prime \, g}{2 \,
\Lambda^2} \, H^\dagger \, \tau_k \, H \, W_k^{\mu \, \nu}
\widetilde B_{\mu\, \nu}  .
\end{eqnarray}
Here, $g^\prime$ is the weak hypercharge gauge coupling,
$B_{\mu\nu}$ is the field strength tensor for the hypercharge
gauge group, $W_k^{\mu \nu}$ is the field strength tensor for the
weak $SU(2)$ gauge group ($k=1,2,3$), $H$ represents the Higgs doublet, and
$\tau_k$ are the Pauli matrices for weak isospin. The operators
$\mathcal{O}_i$ are $CP$ even, and $\widetilde{\mathcal{O}}_j$ are
$CP$ odd. The dual field-strength tensors are defined by
$\widetilde{X}_{\mu \nu} = (1/2) \, \varepsilon_{\mu \nu \alpha
\beta} X^{\alpha \beta}$, for $X=B,W_k$. The corresponding
effective Hamiltonian is
\begin{equation}\label{eq:007}
{\cal H}^{(6)}_{\rm eff}=-{\cal L}^{(6)}_{\rm
eff}=\sum_{i}c_i\,\mathcal{O}_i+\sum_{j}\tilde{c}_j\widetilde{\mathcal{O}}_j\,,
\end{equation}
where $i=(B,\,W,\,BB,\,WW,\,WB)$ and $j=(BB,\,WW,\,WB)$. The
$h\,\gamma\,\gamma$ and $h\,\gamma\, Z$ couplings follow from the
effective Lagrangian (\ref{eq:007}) by making the replacement
$H\to \left(0\,,\left(v+h\right)/\sqrt{2}\right)^T$ in the unitary gauge,
\begin{equation}\label{eq:008}
c_\gamma^{\rm NP}=\left(\frac{4\,\pi
v}{\Lambda}\right)^2\left(c_{WB}-c_{BB}-c_{WW}\right)\,,
\end{equation}
\begin{equation}\label{eq:009}
{\tilde c}_\gamma=\left(\frac{4\,\pi
v}{\Lambda}\right)^2\left(\tilde{c}_{BB}+\tilde{c}_{WW}-\tilde{c}_{WB}\right)\,,
\end{equation}
\begin{eqnarray}\label{eq:010}
c_{1Z}^{\rm NP}&=&\left(\frac{4\,\pi
v}{\Lambda}\right)^2\sin\theta_W\Bigl(c_{BB}\tan\theta_W-c_{WW}\cot\theta_W\nonumber\\
&+&c_{WB}\cot2\,\theta_W\Bigr)\,,
\end{eqnarray}
\begin{equation}\label{eq:011}
c_{2Z}=\left(\frac{2\,\pi
v}{\Lambda}\right)^2\frac{c_B-c_W}{\cos\theta_W}\,,
\end{equation}
\begin{eqnarray}\label{eq:012}
{\tilde c}_Z&=&\left(\frac{4\,\pi
v}{\Lambda}\right)^2\sin\theta_W\Bigl(\tilde{c}_{WW}\cot\theta_W-\tilde{c}_{BB}\tan\theta_W\nonumber \\
&-&\tilde{c}_{WB}\cot2\,\theta_W\Bigr)\,.
\end{eqnarray}
The effective dimensionless couplings $c_B$, $c_W$, $c_{BB}$,
$c_{WW}$, $c_{WB}$, $\tilde{c}_{BB}$, $\tilde{c}_{WW}$, and
$\tilde{c}_{WB}$ could be of order unity based on naive
dimensional analysis~\cite{Manohar:1984np,Georgi:1986np}. If the
theory is valid up to a scale $\Lambda\sim 4\,\pi v$ then it
follows from Eqs.~(\ref{eq:008})--(\ref{eq:012}) that
$c_\gamma^{\rm NP}, \, c_{1Z}^{\rm NP},$ \ ${\tilde c}_\gamma$, \
$c_{2Z}$, \ ${\tilde c}_Z$ can be of the order unity.

On the other hand, values of coupling constants $c_\gamma^{\rm
NP}, \, c_{1Z}^{\rm NP},$ \ ${\tilde c}_\gamma$, \ $c_{2Z}$, \
${\tilde c}_Z$ can be calculated in various models. In particular,
there are models with more than one Higgs doublet which induce
$CP$ violation due to the specific coupling of neutral Higgs
bosons to fermions. We calculate $c_\gamma^{\rm NP}, \,
c_{1Z}^{\rm NP},$ \ ${\tilde c}_\gamma$, \ $c_{2Z}$, \ ${\tilde
c}_Z$ assuming that the couplings of $h$ boson to the fermion
fields, $\psi_f$, are given by the Lagrangian including both
scalar and pseudoscalar parts
\begin{equation}\label{eq:013}
{\cal L}^{hff}=-\sum_{f}\frac{m_f}{v}\,h\,{\bar
\psi_f}\left(1+s_f+i\,p_f\gamma_5\right)\psi_f \,,
\end{equation}
where $s_f$, $p_f$ are real parameters and $s_f=p_f=0$
corresponds to the SM.

Evaluating the fermion contribution to the one-loop $h \to \gamma
\, \gamma$ and $h \to \gamma \, Z$ amplitudes we obtain
\begin{eqnarray}\label{eq:014}
c_\gamma^{\rm NP}&=&\sum_{f}
N_f\,s_f\,Q_f^2\,A_{1/2}^\gamma(\tau_f)\nonumber \\
&\approx& 1.84 s_t-\left(3 s_b+2 s_c+2
s_\tau\right) \times 10^{-2}\nonumber \\
&+&i\, 2\left(2 s_b+s_c+ s_\tau\right) \times 10^{-2} \,,
\end{eqnarray}
\begin{eqnarray}\label{eq:015}
{\tilde c}_\gamma&=&-2\,\sum_{f}
N_f\,p_f\,Q_f^2\,\tau_f\,f(\tau_f)\nonumber \\
&\approx& 2.79 p_t+\left(3 p_b+2 p_c+2
p_\tau\right) \times 10^{-2}\nonumber \\
&-&i\, 2\left(2 p_b+p_c+ p_\tau\right) \times 10^{-2} \,,
\end{eqnarray}
\begin{eqnarray}\label{eq:016}
c_{1Z}^{\rm NP}&=&-\sum_{f}
N_f\,s_f\,Q_f\,g_f\,A_{1/2}^Z(\tau_f\,,\lambda_f) \nonumber \\
&\approx& 0.3253 s_t-\left(8.2 s_b+1.2 s_c+0.2
s_\tau\right) \times  10^{-3}\nonumber \\
&+&i\left(4.8 s_b+0.5 s_c+ 0.1 s_\tau\right) \times 10^{-3}\,,
\end{eqnarray}
\begin{eqnarray}\label{eq:017}
{\tilde c}_Z&=&-\sum_{f}
N_f\,p_f\,Q_f\,g_f\,I_2(\tau_f\,,\lambda_f) \nonumber \\
&\approx& -0.4939 p_t+\left(9.6 p_b+1.3 p_c+0.3
p_\tau\right) \times 10^{-3}\nonumber \\
&-&i\left(4.9 p_b+0.5 p_c+ 0.1 p_\tau\right) \times 10^{-3}\, ,
\end{eqnarray}
where one-loop functions $f(\tau_f), \, I_2(\tau_f\,,\lambda_f) $ are specified in the Appendix~\ref{sec:appendix}, and their arguments $\tau_f, \, \lambda_f$ are defined after Eq.~(\ref{eq:005}).

In obtaining the numerical values in
(\ref{eq:014})--(\ref{eq:017}) we have taken into account dominant
contributions from the charm, bottom, top quarks and $\tau$
lepton, in particularly, the charm, bottom quarks and $\tau$
lepton give rise to the imaginary parts of the couplings in
(\ref{eq:014})--(\ref{eq:017}).

In terms of the parameters $s_f$ and $p_f$ the width of the decay
$h \to f \bar{f}$ is written as
\begin{equation}
\Gamma (h \to f \bar{f})\, = \, \frac{N_f G_F}{4 \sqrt{2} \pi} \, m_f^2 \,
m_h \, \beta_f \left(   (1+s_f)^2 \beta_f^2 \, + \, p_f^2  \right)\,,
 \label{eq:0171}
\end{equation}
where $\beta_f = \sqrt{1- 4m_f^2/m_h^2} $ is velocity of fermion
$f = (\ell, q)$ in the rest frame of $h$. With a good accuracy one
can put $\beta_f = 1$. Note that if one chooses $ (1+s_f)^2 +
p_f^2 = 1$, then the width in Eq.~(\ref{eq:0171}) coincides with
the decay width of the SM Higgs boson.


\section{\label{sec:amplitudes} Amplitudes and angular distributions}

Let us consider the decay of the zero-spin Higgs $h$ boson into a
pair of photons
\begin{equation}\label{eq:018}
h(p)\to \gamma(k_1,\epsilon_1)\,\gamma(k_2,\epsilon_2)\,,
\end{equation}
where $p$ is the four-momentum of $h$ boson, $k_1, \, k_2$ are the
four-momenta of photons and $\epsilon_1, \, \epsilon_2$ are the
corresponding polarization four-vectors. In the rest frame of $h$,
the amplitude of this decay can be written in the form
\begin{equation}\label{eq:019}
{\cal A}(h\to 2\,\gamma)=\frac{e^2
m_h^2}{16\,\pi^2\,v}\left(c_\gamma(\vec{e}_1^{\,*} \,
\vec{e}_2^{\,*}) +{\tilde c}_\gamma(\hat{\vec{k}} \,
[\vec{e}_1^{\,*}\times \vec{e}_2^{\,*}]) \right)\,,
\end{equation}
where $m_h$ is the mass of $h$ boson. The polarization vectors are
chosen in the form $\epsilon_1 = (0, \, \vec{e}_1)$, \
$\epsilon_2 = (0, \, \vec{e}_2)$, where $\vec{e}_1 \, \vec{k} =
\vec{e}_2 \, \vec{k} =0 $, $\vec{k}$ \ is the three-momentum of
one of the photons and $\hat{\vec{k}}\equiv \vec{k}/|\vec{k}|$.

The helicity amplitudes for decay (\ref{eq:018}) are equal to
\begin{equation}\label{eq:020}
H_\pm = - \frac{e^2 m_h^2}{16\,\pi^2\,v} \left(c_\gamma \pm i \,
\tilde c_\gamma  \right).
\end{equation}
The decay width of $h \to 2 \gamma$ is
\begin{equation}\label{eq:021}
\Gamma (h \to 2 \gamma) = \frac{1}{32 \, \pi m_h} \left( |H_+|^2 +
|H_-|^2 \right).
\end{equation}

The polarization states of a single photon are usually described
through the density matrix $\rho^{(\gamma)}$. For the process
(\ref{eq:018}), one can write the two-photon density matrix
following Ref.~\cite{Bernstein:1960} as follows:
\begin{eqnarray}
&&\rho^{(\gamma \gamma)} =\frac{1}{4} \left( 1 \otimes 1 -
\sigma_3 \otimes \sigma_3 + \xi_1 \left( \sigma_1 \otimes \sigma_2
- \sigma_2 \otimes \sigma_1   \right) \right.
\nonumber \\
&  + & \left. \xi_2 \left( \sigma_3 \otimes 1  - 1 \otimes
\sigma_3  \right) - \xi_3 \left( \sigma_1 \otimes \sigma_1 +
\sigma_2 \otimes \sigma_2 \right) \right) \, , \label{eq:022}
\end{eqnarray}
where $\vec{\sigma} =( \sigma_1, \, \sigma_2, \, \sigma_3)$ are the Pauli matrices, $1$ is $2 \times 2$ unit matrix,
and $\otimes$ means the direct product of two matrices. The reference frame is chosen with the OZ axis along $\hat{\vec{k}}$, and
matrices on the left (right) from symbol $\otimes$ refer to the photon with
momentum $\vec{k}$ ($-\vec{k}$).

In (\ref{eq:022}) the following parameters are introduced
\begin{eqnarray}
&&\xi_1  \, = \, \frac{2 \, {\rm Im} \left( H_+ H_-^* \right) }{|H_+|^2 +
|H_-|^2 } \, = \,  \frac{2\,{\rm Re}(c_\gamma
\tilde{c}_\gamma^*)}{|c_\gamma|^2+|\tilde{c}_\gamma|^2}\,,\nonumber \\
&&\xi_2 \, = \, \frac{|H_+|^2 - |H_-|^2  }{|H_+|^2 +
|H_-|^2 } \, = \,  \frac{2\,{\rm Im}(c_\gamma
\tilde{c}_\gamma^*)}{|c_\gamma|^2+|\tilde{c}_\gamma|^2}\,,\label{eq:023}\\
&&\xi_3 \, = \, - \frac{2 \, {\rm Re} \left( H_+ H_-^* \right) }{|H_+|^2 +
|H_-|^2 } \, = \, \frac{|\tilde{c}_\gamma|^2-|c_\gamma|^2}{|c_\gamma|^2+|\tilde{c}_\gamma|^2}\, . \nonumber
\end{eqnarray}
The Stokes parameter $\xi_2$ defines degree of the circular polarization of the photon with momentum $\vec{k}$, it has the meaning
of average photon helicity. Parameters $\xi_1, \, \xi_3$ define  correlation of linear polarizations of two photons (in particular, for $\xi_1=0, \, \xi_3 = -1$ the linear polarizations are parallel, while for $\xi_1=0, \, \xi_3 = 1$ they are orthogonal).

Next we come to the decay of $h$ to $\gamma$ and $Z$ boson
\begin{equation}\label{eq:024}
h(p)\to \gamma(k_1,\epsilon_1)\, Z(k_2,\epsilon_2)\,,
\end{equation}
where $k_1, \, (k_2)$ is the four-momentum of photon ($Z$ boson),
$\epsilon_1, \, (\epsilon_2)$ is polarization  vector of the
photon ($Z$ boson).

The helicity amplitudes for the decay (\ref{eq:024}) are
\begin{equation}\label{eq:025}
H_\pm = - \frac{e g m_h^2}{16\,\pi^2\,v} \left(1 -
\frac{m_Z^2}{m_h^2} \right) \left(c_{1Z} + c_{2Z} \pm i \, \tilde
c_Z \right),
\end{equation}
with the decay width
\begin{equation}\label{eq:026}
\Gamma (h \to \gamma Z) = \frac{1}{16 \, \pi m_h} \left(1 -
\frac{m_Z^2}{m_h^2} \right) \left( |H_+|^2 + |H_-|^2 \right)\,,
\end{equation}
where $m_Z$ is the $Z$ boson mass.

From  definitions (\ref{eq:023}) we find the polarization
parameters
\begin{eqnarray}
&&\xi_1=-\frac{2\,{\rm Im}(A_{\|}\,A_{\perp}^*)}{|A_{\|}|^2+|A_{\perp }|^2}\,,\nonumber \\
&&\xi_2=\frac{2\,{\rm
Re}(A_{\|}\,A_{\perp}^*)}{|A_{\|}|^2+|A_{\perp }|^2}\,,\label{eq:027}\\
&&\xi_3=\frac{|A_{\perp }|^2-|A_{\|}|^2}{|A_{\|}|^2+|A_{\perp
}|^2}\,,\nonumber
\end{eqnarray}
where $H_\pm$ from Eq.~(\ref{eq:025}) for further convenience are
replaced by the amplitudes $A_{\|}=(H_++H_-)/\sqrt{2}$ and
$A_{\perp}=(H_+-H_-)/\sqrt{2}$ corresponding to linearly polarized
final states.

Numerical values of parameters $\xi_1, \, \xi_2, \, \xi_3$ will be discussed in Sec.~\ref{sec:results}.

In the decay (\ref{eq:024}), due to the zero-spin nature of the
Higgs boson, the photon and $Z$ boson have equal helicities. This
allows for measurement of the photon circular polarization through
the decay $h \to \gamma \, Z \to \gamma \,f \bar{f}$~\cite{footnote2}.
Indeed, we derive the following angular distribution of the process in the polar angle $\theta$ between the momentum of the fermion $f$ in the $Z$ boson rest frame and
the direction of the $Z$ boson motion in the $h$ boson rest frame,
\begin{eqnarray}
\frac{1}{\Gamma}\frac{d\Gamma(h\to \gamma \, Z \to \gamma \, f
\bar{f})}{d\cos\theta }&=&\frac{3}{8}\Bigl(1+\cos^2\theta \nonumber \\
&-&2\,A^{(f)}\,\xi_2\,\cos\theta \Bigr)\,,\label{eq:028}
\end{eqnarray}
where
\begin{equation}\label{eq:029}
A^{(f)}\equiv \frac{2\,g_V^{f}
g_A^{f}}{(g_V^{f})^2+(g_A^{f})^2}\,.
\end{equation}
The vector $g_V^{f}$ and axial-vector $g_A^{f}$ constants are
\begin{equation}\label{eq:030}
g_V^f\equiv t_{3L,\,f}-2\,Q_f\sin^2\theta_W\,, \quad g_A^f\equiv
t_{3L,\,f}\,.
\end{equation}
Measurement of the forward-backward asymmetry $A_{\rm FB}$
relative to the direction of $Z$ boson motion in the $h$ boson
rest frame for the $f$ fermions produced in decay (\ref{eq:001}),
\begin{equation}\label{eq:031}
A_{\rm FB}\equiv\frac{F-B} {F+B}\,,
\end{equation}
where
\begin{equation}
F\equiv
\int_0^1\frac{1}{\Gamma}\frac{d\Gamma}{d\cos\theta}\,d\cos\theta \,, \quad \;
B\equiv
\int_{-1}^0\frac{1}{\Gamma}\frac{d\Gamma}{d\cos\theta}\,d\cos\theta
\,,\nonumber
\end{equation}
which is
\begin{equation}\label{eq:032}
A_{\rm FB}=-\frac{3}{4}\,A^{(f)}\,\xi_2\,,
\end{equation}
allows one to find $\xi_2$.

Note that $A^{(\mu)}$ for the decay
\begin{equation}\label{eq:033}
h\to \gamma\, Z \to \gamma\, \mu^-\mu^+
\end{equation}
is $0.142\pm 0.015$ \cite{PDG:2012}, therefore in view of the condition  $|\xi_2| \le 1$, the absolute value of the asymmetry for this decay is not larger than $0.11$. At the same time for the decay channel
\begin{equation}\label{eq:034}
h\to \gamma\, Z \to \gamma\, b\, \bar b
\end{equation}
($A^{(b)}=0.923\pm 0.020$ \cite{PDG:2012}), the absolute value of
$A_{\rm FB}$ can be much larger, namely, as large as $0.69$.

Consider now feasibility to measure the distribution
(\ref{eq:028}) at the LHC after its upgrade to higher luminosity
and energy $\sqrt{s}=14$ TeV. Taking into account various
mechanisms of Higgs boson production in $pp$ collisions, the
inclusive cross section is  \ $\sigma = 57.0163$
pb~\cite{Dittmaier:2011}. Then the cross section for the process
$p \, p \to h \, X \to \gamma \, Z \, X \to \gamma \, \ell^+
\ell^- \, X$ in the SM is
\begin{equation}
\sigma \times {\rm BR}(h \to \gamma Z) \,  {\rm BR}(Z \to \ell^+
\ell^-) = 6.24 \; {\rm fb}\, , \label{eq:0341}
\end{equation}
where $\ell = e, \, \mu$ and the branching fractions are taken
from Refs.~\cite{PDG:2012,Heinemeyer:2013}. In order to observe
the forward-backward asymmetry $A_{{\rm FB}}$ for maximal
value $|\xi_2| = 1$ at a $3 \, \sigma$ level, the number of events should be bigger than 734. This number of events can be obtained, with ideal detector, with  integrated luminosity about 120 fb$^{-1}$.

Let us discuss a possibility to determine the polarization
parameters $\xi_1$ and $\xi_3$. For this one can study the process
\begin{equation}  \label{eq:035}
h \to \gamma^* \, Z \, \to \ell^+ \ell^- \, Z
\end{equation}
with the decay $Z \to \bar{f} f$ on mass shell. For the process  (\ref{eq:035}) we obtain the distribution over the dilepton invariant mass
squared $q^2$ and azimuthal angle $\phi$ between the decay planes
of $\gamma^* \to \ell^+ \ell^-$ and $Z \to \bar{f} f$ in the $h$
rest frame:
\begin{eqnarray}
&&\frac{d\Gamma (h\to \ell^+\ell^-
Z)}{dq^2\,d\phi}/\frac{d\Gamma}{dq^2}=\frac{1}{2\,\pi}\Bigl(1-\frac{1}{4}\left(1-F_L(q^2)
\right)\nonumber \\
&\times&\left(\xi_3(q^2)\cos2\phi+\xi_1(q^2)\sin 2\phi
\right)\Bigr) \,.\label{eq:036}
\end{eqnarray}
Here
\begin{equation}\label{eq:037}
F_L(q^2)\equiv\frac{|A_{0 }(q^2)|^2}{|A_{0 }(q^2)|^2+|A_{\|
}(q^2)|^2+|A_{\perp }(q^2)|^2}
\end{equation}
is the fraction of longitudinal polarization of virtual photon,
and the amplitudes are defined as
\begin{eqnarray}
A_{0 }(q^2) &=&
\frac{e\,g}{16\,\pi^2\,v}\sqrt{\frac{q^2}{m_Z^2}}\Bigl(2\,c_{1Z}\,m_Z^2\nonumber \\
&+&c_{2Z}\left(m_h^2-q^2+m_Z^2 \right)\Bigr)\,,\label{eq:038}
\end{eqnarray}
\begin{eqnarray}
A_{\|}(q^2) &=&
-\frac{e\,g}{8\,\sqrt{2}\,\pi^2\,v}\Bigl(c_{1Z}\left(m_h^2-q^2-m_Z^2 \right) \nonumber \\
&+&c_{2Z}\left(m_h^2+q^2-m_Z^2\right)\Bigr)\,,\label{eq:039}
\end{eqnarray}
\begin{equation} \label{eq:040}
A_{\perp}(q^2) = -i\frac{e\,g}{8\,\sqrt{2}\,\pi^2\,v}{\tilde
c_Z}\sqrt{\lambda(m_h^2,q^2,m_Z^2)}\,,
\end{equation}
with $\lambda(a,b,c)\equiv a^2+b^2+c^2-2\left(a b+a c+b c \right)$
and the distribution over the invariant mass squared reads
\begin{eqnarray}
\frac{d\Gamma}{d q^2}&=&\frac{\alpha_{\rm
em}\,\sqrt{\lambda(m_h^2,q^2,m_Z^2)}}{48\,\pi^2\,m_h^3\,q^2}
\Bigl(\left|A_0(q^2)\right|^2+\left|A_\|(q^2)\right|^2\nonumber \\
&+&\left|A_\perp(q^2)\right|^2\Bigr) \,,\label{eq:041}
\end{eqnarray}
where $\alpha_{\rm em}=e^2/(4\pi)$ is the electromagnetic
fine-structure constant.
The $q^2$-dependent quantities $\xi_1(q^2)$ and $\xi_3 (q^2)$ can
be obtained from Eqs.~(\ref{eq:027}) in which the amplitudes $A_\|
\, (A_\perp)$ are substituted by the $q^2$-dependent amplitudes
$A_\| (q^2) \, (A_\perp (q^2) )$. In derivation of (\ref{eq:036})
we assumed that leptons are massless.

In expressions (\ref{eq:038})--(\ref{eq:040}) we did not take into
account additional two-fermion current operators of dimension 6
\cite{Buchmuller:1986np,Grzadkowski:2010jh} in the effective
Hamiltonian (\ref{eq:007}) and the process $h \to Z^* \, Z \, \to
\ell^+ \ell^- \, Z $. Both these mechanisms contribute at tree
level to the decay $h \to \ell^+ \ell^- \, Z$.

From (\ref{eq:036}) one can approximately find
$\xi_1 $ and $\xi_3 $ in the decay $h \to \gamma \, Z$. Neglecting the amplitude (\ref{eq:038}) for longitudinally polarized photon
$|A_0(q^2)|^2 \sim q^2$, and $q^2$-dependence
of the transverse amplitudes, i.e. substituting $A_{\|}(q^2) \approx
A_{\|}(0)$ and $A_{\perp}(q^2) \approx A_{\perp}(0)$, we
obtain the distribution over the azimuthal angle
\begin{eqnarray}
&&\frac{d\Gamma (h\to \ell^+\ell^- Z)}
{d\phi}\approx\left(\frac{\alpha_{\rm
em}}{3\,\pi}\log\frac{q^2_{\rm max}}{q^2_{\rm
min}}\right)\Gamma(h\to\gamma\,Z)\nonumber \\
&\times&\frac{1}{2\,\pi}\left(1-\frac{1}{4}\left(\xi_3\cos2\phi+\xi_1\sin
2\phi \right)\right) \,.\label{eq:042}
\end{eqnarray}

The lower integration limit $q^2_{\rm min}$ is determined by
possibilities of detectors, in particular, to provide sufficient
$\phi$ resolution to separate $\sin 2 \phi$ and $\cos 2 \phi$
terms in the distribution (\ref{eq:042}).  In this connection we
should mention recent measurements of the $B^0 \to K^{*0} \, e^+
e^-$ branching fraction \cite{LHCb:2013}, in which the LHCb
detector allowed selection of the lower value of dilepton invariant
mass equal to 30 MeV.

Theoretical accuracy of Eq.~(\ref{eq:042}) improves with the decreasing value of $q^2_{\rm max}$, since contribution of the competing
mechanism $h \to Z^* \, Z \, \to \ell^+ \ell^- \, Z $ diminishes
for $q^2_{\rm max} \ll m_Z^2 $.
Consider for example production of the $e^+ e^-$ pair in the
process $h \to e^+e^- \, Z $ with dilepton invariant mass from
30 MeV to 1000 MeV.
Our calculation including both $ h \to \gamma^* \, Z \, \to e^+ e^- \, Z$ and $h \to Z^* \, Z \, \to e^+ e^- \, Z $ amplitudes shows that theoretical error in $\xi_1, \, \xi_3$, which arises when neglecting the $h \to Z^* \, Z \, \to e^+ e^- \, Z $ mechanism, amounts to 20\% in the SM (in which $\xi_1^{SM}=0, \, \xi_3^{SM}=-1$), and 10\% in the effective Hamiltonian approach (\ref{eq:007}) [the choice of coefficients (\ref{eq:008})-(\ref{eq:012}) is discussed in Sec.~\ref{sec:results}].

Of course, the process $ h \to \gamma^* \, Z \, \to e^+ e^- \, Z$ is rare. Let us make an estimate of its observability at the LHC energy $\sqrt{s}=14$ TeV. Using (\ref{eq:042}) and choosing the Higgs production inclusive cross section \ $\sigma = 57.0163$ pb~\cite{Dittmaier:2011} we calculate the SM cross section for the $p \, p \to h \, X \to \gamma^* \, Z \, X \to e^+  e^- \, Z \, X$ in the interval of dilepton invariant mass from $30$ MeV to $1000$ MeV,
\begin{equation}
\sigma \, \times \frac{\Gamma(h \to e^+ e^- Z)|_{30<m_{ee}<1000 \; {\rm MeV }  }}{\Gamma(h \to {\rm
all})} = 0.5 \; {\rm fb} \,. \label{eq:0421}
\end{equation}
When detecting $Z$ boson via $Z \to e^+ e^-$ and $Z \to \mu^+ \mu^-$
channels the cross section (\ref{eq:0421}) is reduced by factor 0.067,
and for the integrated luminosity of 100 fb$^{-1}$ we can expect
about 3 events. This number is too small and a higher integrated luminosity will be needed to observe the decay $h \to \gamma^* \, Z \, \to \ell^+ \ell^- \, Z$ and analyze its angular distribution.


\section{\label{sec:results} Results of calculation and discussion}

First we note that in the SM the polarization parameters are
$\xi_1^{SM} = \xi_2^{SM} =0$ and $\xi_3^{SM} = -1$. Any deviations
of the measured values of $\xi_i$ from $\xi_i^{SM}$ \ ($i=1,2,3$)
will indicate presence of effects beyond the SM.

In order to estimate magnitude of effects of NP, we consider (i) the
approach in which NP is expressed through dimension-6 operators described by effective Hamiltonian (\ref{eq:007}), and (ii) the
model (\ref{eq:013}) with the scalar and pseudoscalar couplings of
fermions to the Higgs boson.

In the approach (\ref{eq:007}) we take for definiteness
$c_{B}$=$c_W$=1, $c_{WB}$=$c_{BB}$=$c_{WW}$=1,
$\tilde{c}_{WB}$=$\tilde{c}_{BB}$=$\tilde{c}_{WW}$=1.
Choosing the scale $\Lambda = 4 \pi v \approx 3.1$ TeV we obtain for the $h \to
\gamma \gamma$ decay
\begin{eqnarray}
\xi_1 &=& -0.259, \quad \xi_2=0.003, \quad  \xi_3=-0.966, \nonumber \\
\mu_{\gamma\gamma} & \equiv & \frac{\Gamma(h\to \gamma\,
\gamma)}{\Gamma^{\rm SM}(h\to \gamma\, \gamma)} = 1.35 \,
,\label{eq:043}
\end{eqnarray}
and for $h \to \gamma Z$ decay
\begin{eqnarray}
\xi_1 & = & -0.107, \quad \xi_2=0.0001, \quad \xi_3=-0.994, \nonumber \\
\mu_{\gamma Z} & \equiv & \frac{\Gamma(h\to \gamma\,
Z)}{\Gamma^{\rm SM}(h\to \gamma\, Z} = 1.12 \, . \label{eq:044}
\end{eqnarray}

For another scale $\Lambda = 2$ TeV, for the $h \to \gamma \gamma$
decay, we obtain
\begin{eqnarray}
\xi_1 & = & -0.497, \quad \xi_2=0.004, \quad \xi_3=-0.868, \nonumber \\
\mu_{\gamma\gamma} &=& 1.99 \, ,\label{eq:045}
\end{eqnarray}
and for $h \to \gamma Z$ decay
\begin{eqnarray}
\xi_1 & = & -0.236, \quad \xi_2=0.0002, \quad \xi_3=-0.972,\nonumber \\
 \mu_{\gamma Z} &=& 1.31 \, . \label{eq:046}
\end{eqnarray}

For the ratio $\mu_{\gamma \gamma}$ our calculation with the scale $\Lambda=4 \pi v$ better agrees with the ATLAS data~\cite{ATLAS:2013_CONF034} for $h \to \gamma \gamma $ than calculation with $\Lambda = 2$ TeV.

In the model with scalar and pseudoscalar couplings of fermions to
the Higgs boson (\ref{eq:013}) we choose the parameters
\begin{eqnarray}
&&p_t=p_b=p_c=p_\tau=\pm\,1/\sqrt{2}\,,\nonumber \\
&&s_t=s_b=s_c=s_\tau=1/\sqrt{2}-1 \label{eq:047}
\end{eqnarray}
satisfying normalization $(1+s_f)^2+p_f^2=1$ discussed in
Sec.~\ref{sec:formalism}.

As a result, for the decay $h\to \gamma\,\gamma$ we find
\begin{eqnarray}
\xi_1 &=& \mp0.528, \quad \xi_2=\mp0.010, \quad \xi_3=-0.849, \nonumber \\
\mu_{\gamma\gamma} &=& 1.26 \label{eq:048}
\end{eqnarray}
and for decay $h\to \gamma\, Z$
\begin{eqnarray}
\xi_1 &=& \pm 0.121, \quad \xi_2=\mp 0.001, \quad \xi_3=-0.993, \nonumber \\
\mu_{\gamma Z } &=& 1.04 .\label{eq:049}
\end{eqnarray}

In addition, the $h \to f \bar{f}$ decay width calculated with $s_f, \, p_f$ in (\ref{eq:047})
coincides with the SM decay width and agrees with the CMS data~\cite{CMS:2013_HIG005}
for $h \to \tau^+ \tau^-$ and $h \to b \, \bar{b}$ decays,
\begin{eqnarray}
\mu_{\tau \tau } & \equiv & \frac{\Gamma(h\to \tau^+\,
\tau^-)}{\Gamma^{\rm SM}(h\to \tau^+\, \tau^-)} = 1.10 \pm 0.41 \, ,
\nonumber \\
\mu_{b \, b } & \equiv & \frac{\Gamma(h\to b \,
\bar{b})}{\Gamma^{\rm SM}(h\to b\, \bar{b})} = 1.15 \pm 0.62 \, .
\label{eq:0471}
\end{eqnarray}

At the same time the channel $h \to c \, \bar{c}$ is not measured
yet. Thus the $h \to c \, \bar{c}$ width, in general, may differ
from the SM prediction, and consequently the constraint
$(1+s_c)^2+p_c^2=1$ for the charm quark may not hold. We can make
an assumption that $\Gamma(h \to c \, \bar{c}) \, \leq \,
\Gamma(h\to b \, \bar{b})$. Combining this inequality with
Eqs.~(\ref{eq:0171}) and (\ref{eq:0471}) we find
\begin{equation}
(1+s_c)^2+p_c^2 \, \leq \, \mu_{bb} \times \frac{\Gamma^{\rm SM}(h\to
b\, \bar{b})}{\Gamma^{\rm SM}(h\to c\, \bar{c})} \,.
\label{eq:0472}
\end{equation}
Taking the central values of $\mu_{bb}$ and the
widths from~\cite{Heinemeyer:2013} (Table~1 therein) we obtain the
following constraint for the $h \, c \, \bar{c}$ couplings: \,
$(1+s_c)^2+p_c^2 \, \leq \, 22.8$.

To estimate maximal values of polarization parameter $\xi_2$ in the channel $h \to \gamma \, Z$ let us take $s_c, \, p_c $ satisfying $(1+s_c)^2+p_c^2 \, = \, 22.8$, although the latter equality does not fix $s_c, \, p_c$ uniquely. In addition, put $s_f = p_f =0$ for $f \ne c$.
Then calculation using (\ref{eq:016}) and (\ref{eq:017}) gives values of $\xi_2$ which do not exceed $8.6 \times 10^{-4}$. It is seen that even for such a radical modification of the Higgs couplings to the charm quarks, the parameter $\xi_2$ remains very small.

Thus the existing data on the Higgs boson decay to the $\tau^+
\tau^-$ and $b \, \bar{b}$ pairs and a reasonable assumption on
the upper bound of the decay width to the charm quarks lead to
conclusion that the rescattering effects on the one-loop level
result in values of $\xi_2$ in the $h \to \gamma Z$ decay about
$10^{-3}$ or smaller.

It would be of interest to check in the experimental analysis of
the distribution (\ref{eq:028}) whether the parameter $\xi_2$ is
very small indeed. If the analysis yielded sizable values of
$\xi_2$, this would mean the presence of additional sources of
non-Hermiticity of effective Lagrangian. The latter may arise, for
example, due to the breaking of Hermiticity in an underlying
(fundamental) theory at very small distances. Note, that similar
aspects have been discussed in~\cite{Ginzburg:2001} for the
process $\gamma \gamma \to h$, where the authors calculated
various asymmetries as functions of complex coefficients
$c_{\gamma}, \, {\tilde c_\gamma}$ in Eq.~(\ref{eq:002}). Since
the requirement of Hermiticity is one of the conditions in the
proof of the $CPT$ theorem~\cite{Streater:1964}, measurement of
the photon circular polarization in the decay $h \to \gamma Z \to
\gamma \bar{f} f$ through the forward-backward asymmetry $A_{\rm
FB}$ can be useful for testing $CPT$ symmetry.

The parameters $\xi_1$ and $\xi_3$ carry information on the $CP$
properties of the Higgs boson.
Besides, $\xi_1$ is $CP$-odd and $T$-odd observable and, in the absence of final-state interaction between the leptons and fermions, a nonzero value of $\xi_1$ will point to the violation of $T$ invariance.


\section{\label{sec:conclusions} Conclusions}

In this paper polarization properties of the $\gamma \gamma$ and
$\gamma Z$ states in the decays $h \to \gamma \gamma$ and $h \to
\gamma Z$ of recently discovered scalar boson have been
considered. We have chosen effective Lagrangian, describing $h
\gamma \gamma $ and $h \gamma Z$ interactions with $CP$-even and
$CP$-odd parts. This allowed for calculation of polarization
parameters $\xi_1, \, \xi_2, \, \xi_3$. In the SM these parameters
take on values $\xi_1^{SM} = \xi_2^{SM} =0$, $\xi_3^{SM} = -1$ and
deviations of the measured values of $\xi_i$ from $\xi_i^{SM}$ \
($i=1,2,3$) will point to effects of NP.

The parameter $\xi_2$, which defines the circular polarization of
the photon, can be measured in the  $h \to \gamma \, Z \, \to
\gamma \, f\, \bar{f}$ decay through the forward-backward
asymmetry $A_{\rm FB} \sim \xi_2$ of the fermion $f$. The parameters $\xi_1, \,
\xi_3$, which define correlation of linear polarizations of $\gamma$ and $Z$,
can be extracted from the azimuthal angle distribution in the process
$h \to \gamma^* \, Z \, \to \ell^+ \ell^- \, Z$ with decay $Z \to \bar{f}
f$ on the mass shell.

In numerical estimates of these parameters we included the one-loop
contribution from the SM, and models beyond the
SM. Namely, we applied the approach~\cite{Manohar:2006pl,Buchmuller:1986np,Hagiwara:1993pl,Hagiwara:1993pr,Grzadkowski:2010jh,Grojean:2013} in which NP is described by dimension-6 operators in the fields of the SM,
and model with scalar and pseudoscalar
couplings of fermions to the Higgs boson on the one-loop level.

The value of photon circular polarization turns out to be very small, of the order $10^{-3}$. In general,
nonzero value of $\xi_2$ arises due to presence of the $CP$-even
and $CP$-odd parts in effective Lagrangian ${\cal L}_{\rm eff}^{h\gamma Z}$ and absorptive parts of one-loop diagrams, or
rescattering effects of the type $h \to a \bar{a} \to \gamma \,
Z$, where $a$ are charged particles with masses $m_a \le m_h/2$.
Only leptons and quarks $u, \,d, \, s,\, c,\, b$ satisfy this
condition and hence contribute to
absorptive parts of one-loop diagrams. Contributions from leptons
$e, \, \mu$ and light quarks $u, \, d, \, s$ are negligibly small.
The couplings of $h$ to the $\tau$ lepton and bottom quark are
constrained by recent CMS data on the $h \to \tau^+ \tau^-$ and $h
\to b \, \bar{b}$ \ decays,
and couplings to the charm quark are constrained from an assumption on the upper bound of the $h \to c \, \bar{c}$ \ decay width.

Apart from rescattering effects, in framework of
$CPT$ symmetric models, there are no sources of non-Hermiticity of
${\cal L}_{\rm eff}^{h\gamma Z}$ which could contribute to
parameter $\xi_2$.
If there is a violation of $CPT$ symmetry in an
underlying theory at small distances, then this may give rise to
additional non-Hermiticity effects in ${\cal L}_{\rm eff}^{h\gamma
Z}$ which will change the value of $\xi_2$.  Therefore measurement of
this parameter in the $h \to \gamma \, Z \, \to \gamma \, f\,
\bar{f}$ process would allow one to test the prediction of the SM, and
to search for deviations from the SM, and even possible effects of
$CPT$ violation in an underlying theory.

Nonzero values of parameter $\xi_1$ point to violation of
$CP$ symmetry in the $h \to \gamma \gamma$ and $h \to \gamma Z$
decays. In the chosen models of NP, for the $h \to \gamma Z$
decay, $\xi_1$ appears to be 0.1-0.2. Its experimental
determination can put constraints on models describing physics
beyond the SM.

We also estimated in the SM a feasibility of measurement of the
discussed processes in the $pp$ collisions at the LHC, after its
upgrade to energy $\sqrt{s}=14$ TeV and higher luminosity.
The cross section for the process $p \, p \to h \, X \to \gamma \,
Z \, X \to \gamma \, \ell^+ \ell^- \, X$ \ ($\ell = e, \, \mu$)
turns out to be 6.24 fb. With integrated luminosity about 120
fb$^{-1}$ and ideal detector it may be possible to observe 
the forward-backward asymmetry $A_{\rm FB}$ 
for $|\xi_2| = 1$ at a $3 \, \sigma $ level. 

Here we should mention papers~\cite{Gainer:2012,Campbell:2013},
where possibilities of studying at the LHC the $h \to \gamma \,
\ell^+ \ell^- $ decay via $\gamma \, Z $ channel are considered.
Although observation of the Higgs is difficult in view of the
background which is a few orders of magnitude larger than the
signal and unfavorable kinematics of this
decay~\cite{Campbell:2013}, in these papers optimistic conclusions
are made as for measurement of the branching ratio of the SM Higgs
decay to $\gamma \, Z$ at the 14 TeV LHC with integrated
luminosity of 100 fb$^{-1}$~\cite{Gainer:2012}.

The reaction $p \, p \to h \, X \to \gamma^* \, Z \, X \to e^+
e^- \, Z \, X$ is a more rare process, and our estimate of its
observability is less optimistic. One can expect about 3 events in
the interval of $e^+ e^-$ invariant mass from $30$ MeV to $1000$
MeV if $Z$ boson is detected through the $Z \to e^+ e^-, \, \mu^+
\mu^-$ channels. Clearly an integrated luminosity higher than 100
fb$^{-1}$ will be needed to study the $h \to \gamma^* \, Z \, \to
e^+ e^- \, Z$ process.

In conclusion, we hope that with increasing the integrated
luminosity at the LHC investigation of angular distributions
discussed in the present paper will become possible.


\appendix

\section{\label{sec:appendix} Definition of Loop Functions }

The loop functions for the $W^\pm$ boson ($A_1^{\gamma\,(Z)}$) as
well as the fermion $f$ ($A_{1/2}^{\gamma\,(Z)}$) are defined in
Ref.~\cite{Spira:1998}
\begin{equation}\label{appendixA:001}
A_1^\gamma(\tau)=-\left(2+3\tau+3\tau(2-\tau)f(\tau)\right)\,,
\end{equation}
\begin{equation}\label{appendixA:002}
A_{1/2}^\gamma(\tau)=2\tau\left(1+(1-\tau)f(\tau)\right)\,,
\end{equation}
\begin{eqnarray}
&&A_1^Z(\tau,\lambda)=\cos\theta_W
\Bigl(4\left(3-\tan^2\theta_W\right)I_2(\tau,\lambda)\nonumber \\
&+&\left(\left(1+\frac{2}{\tau}\right)\tan^2\theta_W
-\left(5+\frac{2}{\tau}\right)\right)I_1(\tau,\lambda)\Bigr)\,,\label{appendixA:003}
\end{eqnarray}
\begin{equation}\label{appendixA:004}
A_{1/2}^Z(\tau,\lambda)=I_1(\tau,\lambda)-I_2(\tau,\lambda)\,.
\end{equation}
The functions $I_1$, $I_2$ are given by
\begin{eqnarray}
I_1(\tau,\lambda)&=&\frac{\tau\,\lambda}{2\,(\tau-\lambda)}\Bigl(1+\frac{\tau\,\lambda}{\tau-\lambda}
(f(\tau)-f(\lambda)) \nonumber \\
&+&\frac{2\,\tau}{\tau-\lambda}(g(\tau)-g(\lambda))\Bigr)\,,\label{appendixA:005}
\end{eqnarray}
\begin{equation}\label{appendixA:006}
I_2(\tau,\lambda)=-\frac{\tau\,\lambda}{2\,(\tau-\lambda)}\left(f(\tau)-f(\lambda)\right)\,,
\end{equation}
where the functions $f(\tau)$ and $g(\tau)$ can be expressed as
\begin{eqnarray}\label{appendixA:007}
f(\tau)=\left\{
\begin{array}{ll}  \displaystyle
\arcsin^2\frac{1}{\sqrt{\tau}} & \tau\geq 1 \\
\displaystyle -\frac{1}{4}\left( \log\frac{1+\sqrt{1-\tau}}
{1-\sqrt{1-\tau}}-i\pi \right)^2 \hspace{0.5cm} & \tau<1\,,
\end{array} \right.
\end{eqnarray}
\begin{eqnarray}\label{appendixA:008}
g(\tau)=\left\{
\begin{array}{ll}  \displaystyle
\sqrt{\tau-1}\,\arcsin\frac{1}{\sqrt{\tau}} & \tau\geq 1 \\
\displaystyle \frac{\sqrt{1-\tau}}{2}\left(
\log\frac{1+\sqrt{1-\tau}} {1-\sqrt{1-\tau}}-i\pi \right)
\hspace{0.5cm} & \tau<1\,.
\end{array} \right.
\end{eqnarray}



\begin{thebibliography}{99}
\bibitem{footnote2} Of course for background processes, in which the photon and $Z$ boson helicities can differ, this statement is not valid.
\end{thebibliography}
\end{document}